\documentclass{elsart}

\usepackage{amssymb}

\usepackage{graphicx}
\usepackage{dcolumn}
\usepackage{bm}

\begin{document}

\begin{frontmatter}


\title{Statistical methods applied to composition studies of ultrahigh energy cosmic rays} 

\author[ifgw]{F. Catalani} \ead{catalani@ifi.unicamp.br}
\author[ifgw]{J.A. Chinellato} \ead{chinella@ifi.unicamp.br}
\author[iag,unikar]{V. de Souza} \ead{desouza@ik.fzk.de}
\author[ifgw]{J. Takahashi} \ead{jun@ifi.unicamp.br}
\author[ifgw]{G.M.S. Vasconcelos} \ead{gmagela@ifi.unicamp.br}
\address[ifgw]{Instituto de F\'{\i}sica Gleb Wataghin, Universidade
Estadual de Campinas, Brasil} 
\address[iag]{Instituto de Astronomia,
Geof\'isica e Ci\^encias Atmosf\'ericas \\ Universidade de S\~ao
Paulo, Brasil}
\address[unikar] {Institut f\"ur Experimentelle Kernphysik, Universit\"at Karlsruhe, 76021 Karlsruhe, Germany. }


\begin{abstract}
The mass composition of high energy cosmic rays above $10^{17}$ eV is
a crucial issue to solve some open 
questions in astrophysics such as the acceleration and propagation mechanisms.
Unfortunately, the standard procedures to identify the primary particle of a cosmic ray shower have low 
efficiency mainly due to large fluctuations and limited experimental observables. We present a
statistical method for composition studies based on several measurable features of the longitudinal 
development of the CR shower such as $N_{max}$, $X_{max}$, asymmetry, skewness and kurtosis.  
Principal component analysis (PCA) was used to evaluate the relevance of each parameter in the 
representation of the overall shower features and a linear discriminant analysis (LDA) was used to combine 
the different parameters to maximize the discrimination between different particle showers. 
The new parameter from LDA provides a separation between primary gammas, proton and iron nuclei better than 
the procedures based on $X_{max}$ only. The method proposed here was successfully tested in the energy 
range from $10^{17}$ to $10^{20}$ eV even when limitations of shower track length were included in order 
to simulate the field of view of fluorescence telescopes.
\end{abstract}


\end{frontmatter}

\section{\label{introduction} Introduction}

The most important questions about cosmic rays with energy above $10^{17}$ eV can only be 
solved if the mass composition spectra is known. The complete understanding of the energy
spectra and its features is only possible with the knowledge of the mass composition since 
it stores information about the source and the acceleration mechanisms.
The evolution of the energy spectrum and the explanation of its
possible features (2nd knee, ankle 
and the GZK cutoff) are intrinsically linked to the primary composition because the source 
power (in a Fermi mechanism) is directly proportional to the particle charge. 
Furthermore, the attenuation length of the particles due to the various energy loss 
mechanisms depends on the particle type and thus different mass composition can modulate 
the energy spectrum up to the highest energies. 

Present measurements with different techniques are not conclusive about the mass 
spectrum and its evolution with energy. Currently, composition studies have been done 
using fluorescence telescopes and ground arrays. Fluorescence telescopes are sensitive to
the primary composition by measuring the depth in which the shower has the maximum number 
of particles ($X_{max}$) while ground based experiments usually extract the composition 
information from the muon density or the temporal structure of the signal. However, due to
the large fluctuations in the shower development, the limited number of measurable parameters 
and the uncertainties in the hadronic interaction models \cite{bib:dova,bib:kascade} the 
primary particle identification is very difficult and not possible on a event by event basis.

Several experiments have published mass composition studies. The Haverah Park experiment has 
reported an iron fraction around 66\% in the energy range 0.2-1 EeV \cite{bib:hp:compo} and 
similar results have been measured by the Akeno \cite{bib:akeno:compo} experiment. At energies 
above $10^{19}$ eV, the AGASA experiment \cite{bib:agasa:compo} measured an upper limit of the 
iron fraction of 35\% in the range $10^{19}-10^{19.5}$ eV and 76\% in the range 
$10^{19.5}-10^{20}$ eV. On the other hand, the HiRes Collaboration \cite{bib:hires:compo}, 
measured an unchanging light composition above $10^{18}$ eV and a change from heavy to light 
composition in the range $10^{17}-10^{18}$ eV. Thus, the composition measurement for energies 
above $10^{19.5}$ eV are still inconclusive \cite{bib:Watson}.

In this paper, we present the application of a statistical method to differentiate the primary 
particle based on several features of the longitudinal development of the CR shower rather than 
using only the depth of shower maximum ($X_{max}$).
The longitudinal development of the CR shower can be well measured by fluorescence telescopes 
in operation by the HiRes and Auger Collaborations \cite{bib:Auger}. Our results are based on Monte 
Carlo simulations of the showers.

This article is organized as follows: section \ref{sec:simulation} describes the longitudinal 
profile and the $X_{max}$ parameter that is commonly used for composition studies. 
Section \ref{sec:pca} describes the additional shower parameters proposed in this paper and the 
principal component analysis (PCA) used to evaluate them.  Section \ref{sec:lda} describes 
the linear discriminant analysis (LDA) and the results obtained using the newly defined parameters. 
Section \ref{sec:telescope} shows the results of the new method  when the field of view of 
fluorescence telescopes is taken into account. A discussion of our results and future 
perspectives are summarized in the final section.

\section{\label{sec:simulation} Shower longitudinal profile and the $X_{max}$ parameter}

We have simulated air showers using the CORSIKA version 6.203 \cite{bib:corsika} Monte Carlo 
program with the QGSJET 01 \cite{bib:qgsjet} hadronic interaction model. Particles in the shower were 
followed down to the energy of 0.05 GeV (muon and hadrons) and 50 keV (photons and
electrons). The longitudinal development of the showers was sampled in vertical steps of 
5 g/cm$^2$. The thinning algorithm \cite{bib:hillas,bib:heck} was used
with a thinning factor of $10^{-4}$ and a  
maximum weight of $10^{5}$. Primary photons have been simulated with the pre-shower \cite{bib:preshow} 
effect and . Two thousand shower were simulated for each type of
primary particle and for each energy ranging  
from $10^{17}$ eV to $10^{20}$ eV.

\begin{figure}[]
\centerline{\includegraphics[width=7cm, angle=90]{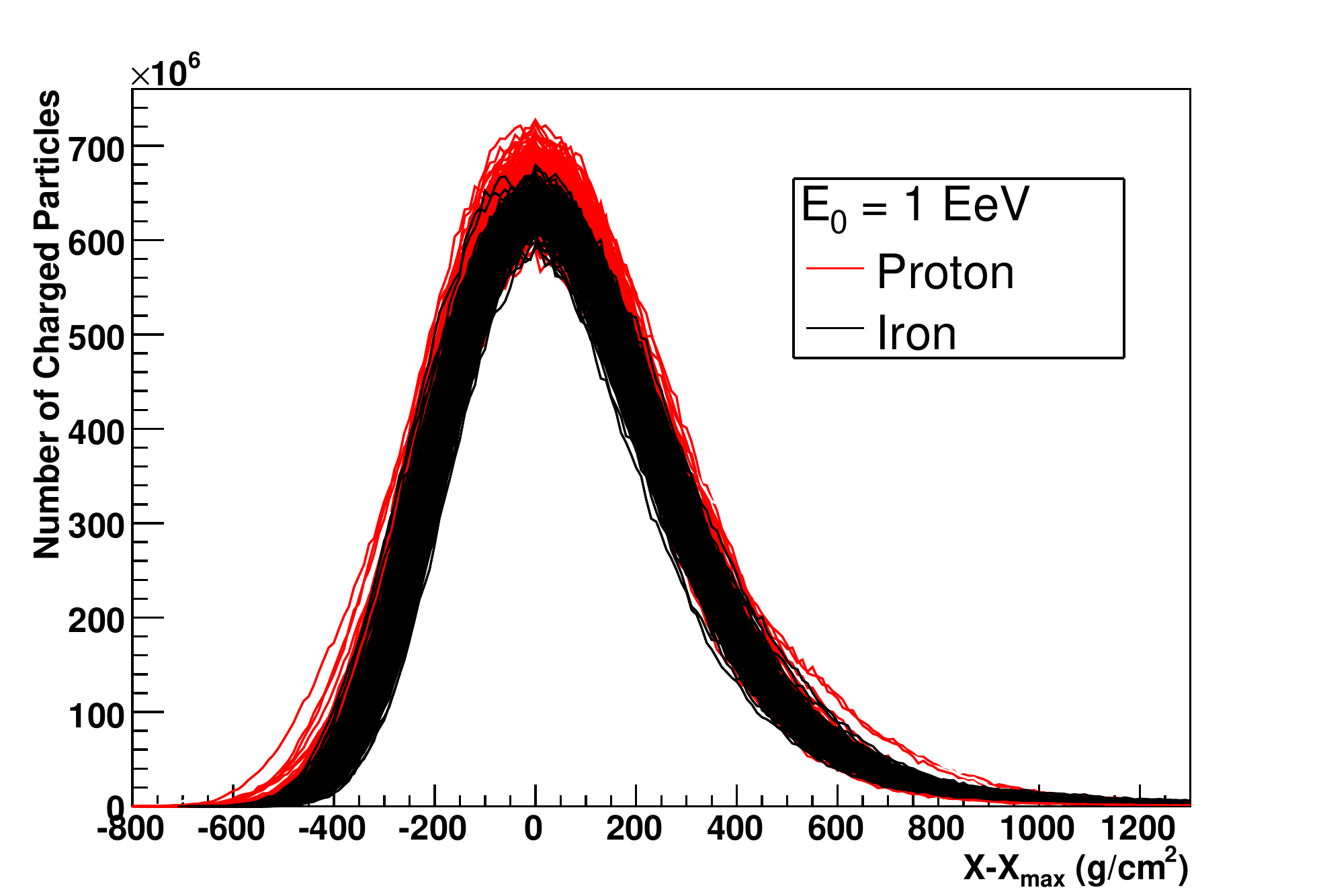}}
\caption{Longitudinal development of proton and iron nuclei
  showers. Each profile was shifted by its $X_{max}$ depth. Figure
  shows 100 proton and 100 iron nuclei showers.}
\label{fig:longitudinal}
\end{figure}

Figure \ref{fig:longitudinal} shows the simulated longitudinal profile of iron and proton showers. 
Shower profiles shown in this figure were shifted by its $X_{max}$ for better 
comparison of the profile shapes.

\begin{figure}[]
\centerline{\includegraphics[width=7cm, angle=90]{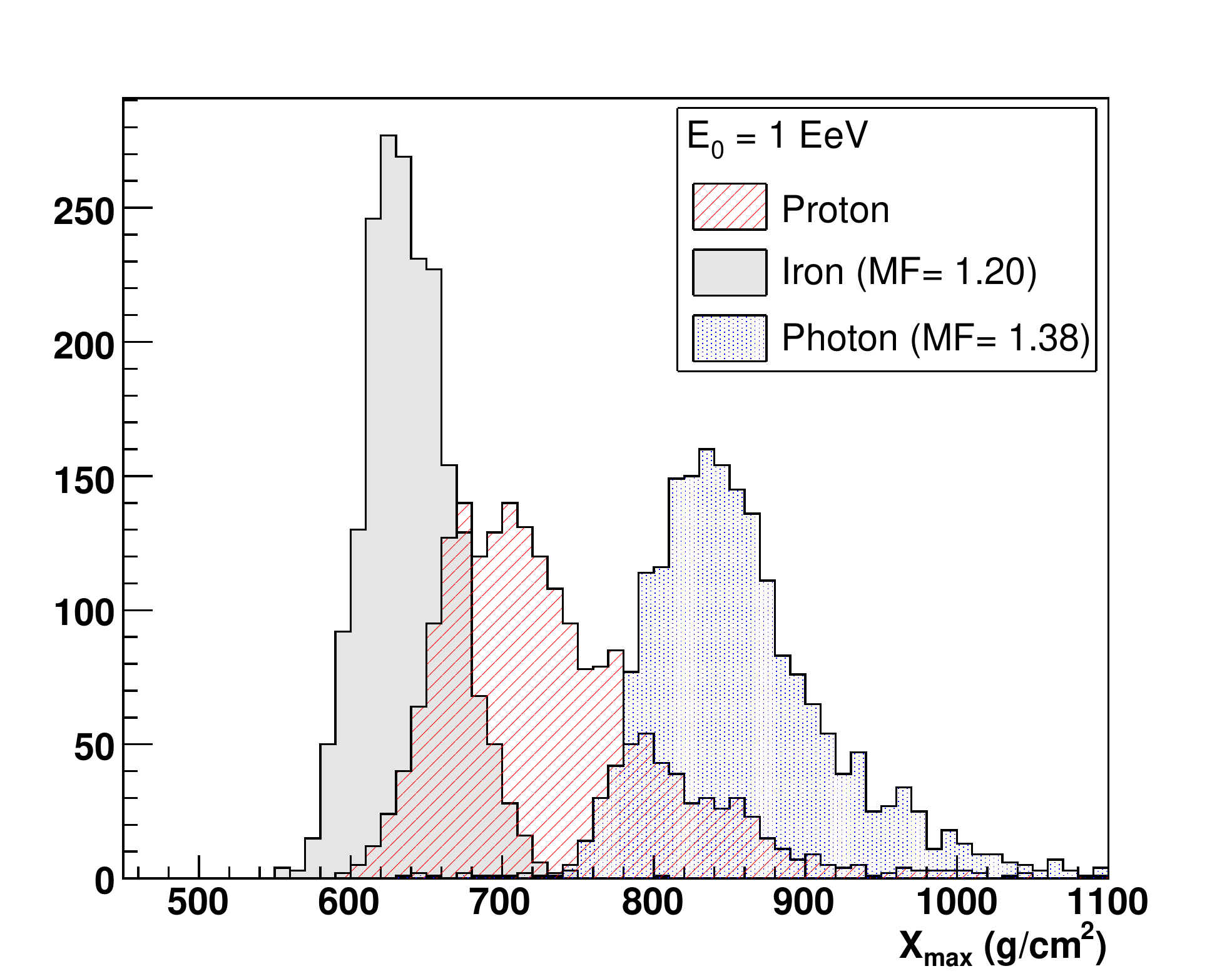}}
\caption{Distribution of $X_{max}$ for proton, iron nuclei and photon primaries.}
\label{fig:xmax}
\end{figure}

\begin{figure}[]
\centerline{\includegraphics[width=7cm, angle=90]{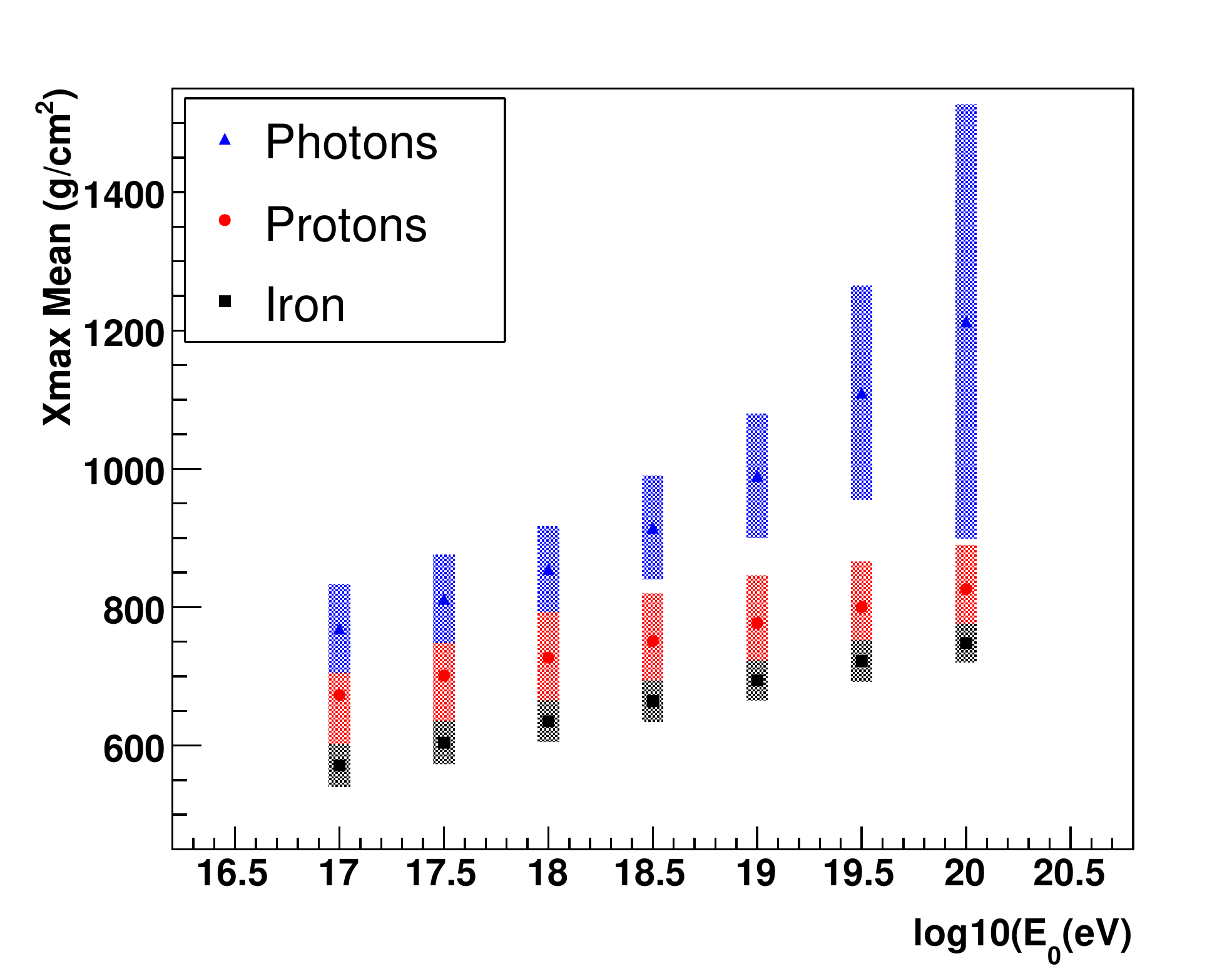}}
\caption{Simulated elongation rate for proton, iron nuclei and photon
  primaries. Error bars represent the RMS of the $X_{max}$ distribution.}
\label{fig:elongation}
\end{figure}

Composition studies done by fluorescence telescopes are usually based on this $X_{max}$ parameter. 
Figure \ref{fig:xmax} shows the $X_{max}$ distribution for primary protons, iron nuclei and photons
for showers simulated with primary energy of $10^{18}$ eV. 
This figure illustrates that the $X_{max}$ parameter can be a good discriminator for gamma showers 
but its capability to differentiate protons from iron nuclei primary particles is quite limited 
due to the large overlap between the two distributions. 
In the energy range between $10^{17}$ eV to $10^{20}$ eV the discrimination capability of 
$X_{max}$ varies as shown in figure \ref{fig:elongation}. The elongation rate shows how the average
$X_{max}$ varies  with energy and is normally used to report the evolution of the composition with 
energy. Figure \ref{fig:elongation} also shows the RMS of the $X_{max}$ distribution as the error bars 
and the clear overlap of the $X_{max}$ parameter for different primary particles indicates that the
discrimination between the different particles becomes more difficult for higher energies.

To quantify the separation capability hence the discrimination between the different primary particle 
distributions,  we are going to use the merit factor (MF) statistical parameter. The merit factor 
between two distributions (A and B) is defined as:

\begin{equation}
MF = \frac{  \bar{A} - \bar{B} }{\sqrt{ \sigma_A^2 + \sigma_B^2}},
\end{equation} 

where $\bar{A}$ and $\bar{B}$ are the distributions averages, and $\sigma_A$ and $\sigma_B$ the respective 
standard deviations.

\begin{figure}[]
\centerline{\includegraphics[width=7cm]{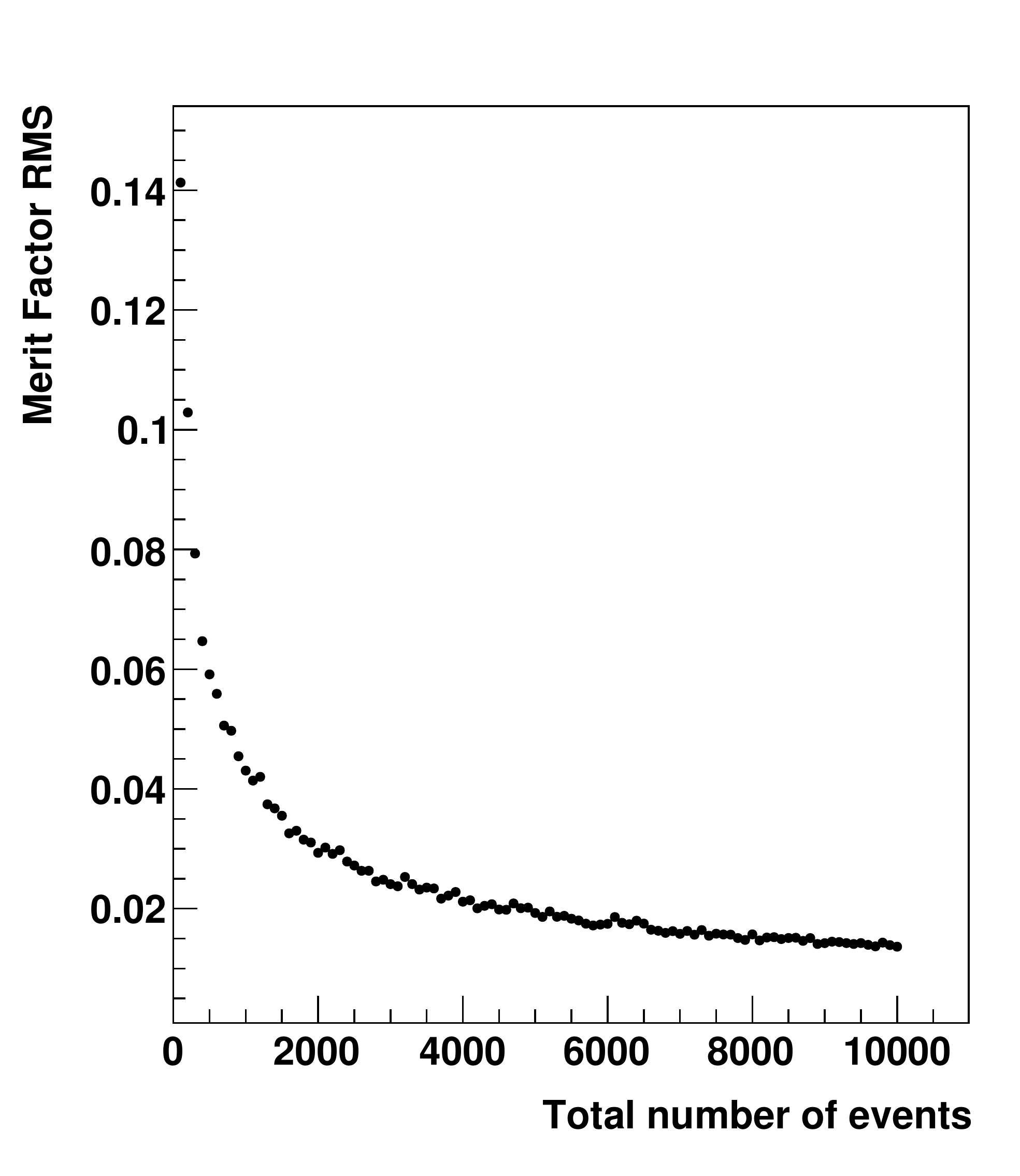}
\includegraphics[width=7cm]{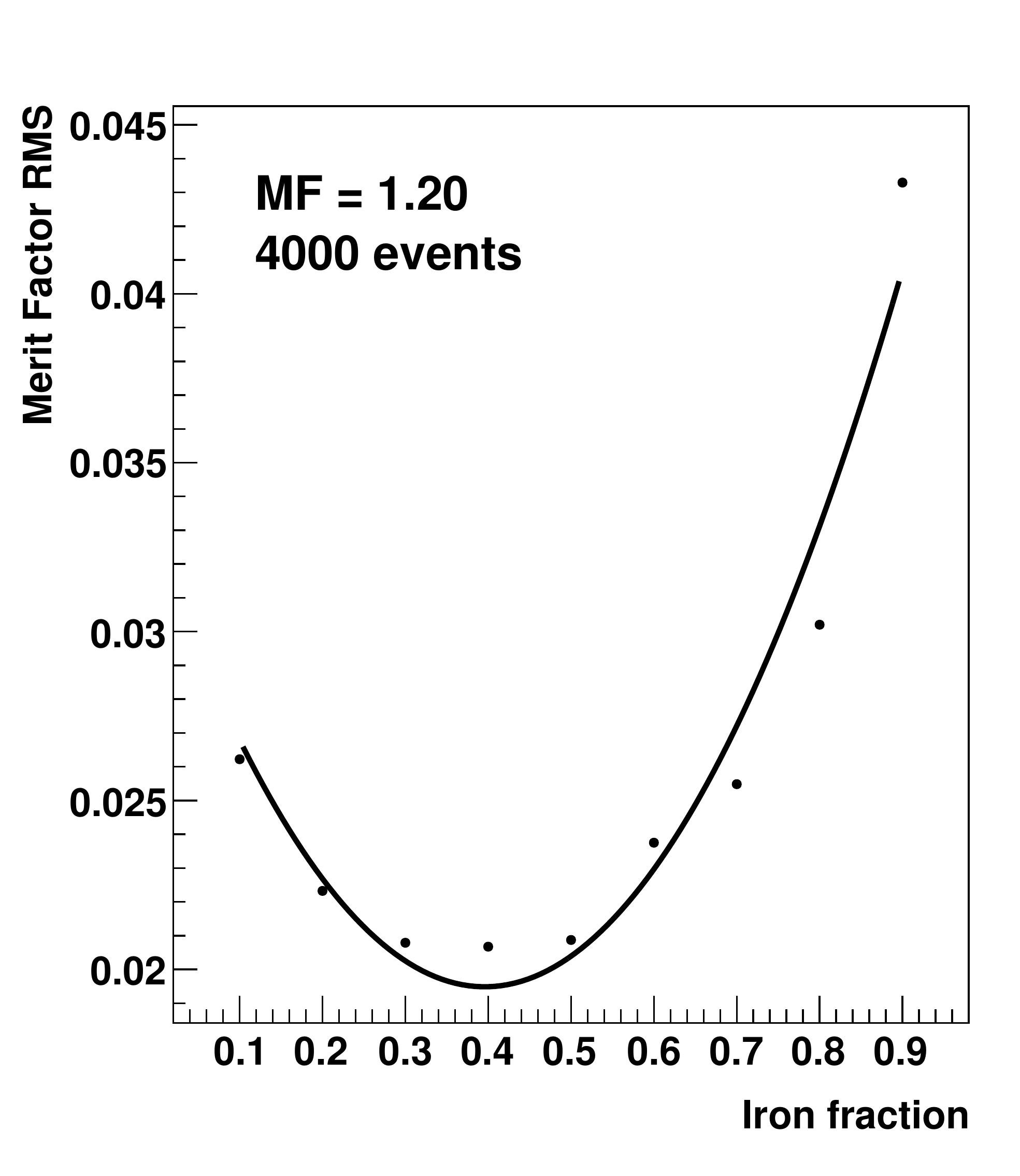}}
\caption{(a) Error dependence of the merit factor with the number of
  events (b) and merit factor error dependence on the relative
  fraction of events between proton showers and iron showers.} 
\label{fig:mf:rms}
\end{figure}

Figure \ref{fig:mf:rms}a shows the merit factor fluctuation as a function of the number of events. 
We have used the $X_{max}$ distribution of primary iron nuclei and protons as shown in 
figure \ref{fig:xmax} for this calculation. Random values were drawn
following those distribution 1000 times and  
for each time a merit factor value was calculated. Following this procedure the fluctuation of the merit
factor as a function of the number of showers could be calculated.
Figure \ref{fig:mf:rms}a shows that the results presented here using the merit factor parameter 
have uncertainties smaller than 3\% since we are always comparing two distributions with 2000 
events each, summing 4000 total events. We have also studied how the uncertainty varies with 
the fraction of iron nuclei in the total number of events. Figure \ref{fig:mf:rms}b shows that 
for any mixture of iron and proton the merit factor uncertainty is smaller than 3\%.

The merit factor achieved for the separation between proton and iron and proton and photon 
using the $X_{max}$ distributions is 1.20 and 1.38, respectively. These values are  going to be 
compared to the merit factor obtained from the new analysis proposed here.

\section{\label{sec:pca} New parameters and the principal component analysis} 

Comparing the shapes of the longitudinal profile shown in figure \ref{fig:longitudinal}, 
it is possible to notice that there are differences between the two different primary particles.
Thus, we have looked for mathematical parameters that express these differences  to
improve the discrimination capacity between the particles.

Several new parameters were tested and only some have shown good separation capabilities and the
most powerful ones seem to be those related to the asymmetry of the longitudinal profiles. 
Below we list some tested parameters and show its discrimination capability by quoting the merit 
factor (MF) between proton and iron distributions.
 
\underline{$X_{max}$}:the atmospheric depth (g/cm$^{2}$) in which the shower has the maximum 
number of particles. It is the most used composition parameter and is calculated in any analysis 
procedure of fluorescence telescopes data. A indirect measurement of the $X_{max}$ can also be done by
water \v{C}erenkov detectors via the rise time
method~\cite{bib:risetime}. $MF =  1.20$.

\vspace{0.3cm}

\underline{$N_{max}$:} the number of particles in the shower at a $X_{max}$. 
  This is also a standard parameter calculated in any fluorescence telescope analysis and 
  it is directly proportional to the shower energy. If the error in the energy reconstruction is large the 
  inclusion of this parameter in the composition study would lead to a dependence with energy 
  that is hard to disentangle. However, the fluorescence telescopes reconstruct the energy with an error
  of about 15\% what we believe is a safe margin since variations in the energy of an EAS of this
  order do not affect the hypothesis about the chemical composition of the primary. $MF = 1.00$

\vspace{0.3cm}

\underline{Asymmetry and Sigma:} in order to measure the asymmetry of
  the distribution  we have fit an asymmetric function to the longitudinal profile defined as:

if ( $X < X_{max}$)

\hspace{3cm} 
$N_{part} = N_{max} \exp{\frac{X-X_{max}}{Sigma^2}} $

if ( $X > X_{max}$)

\hspace{3cm}
$N_{part} = N_{max}\exp{\frac{X-X_{max}}{Asymmetry^2*Sigma^2}}$

$X_{max}$ and $N_{max}$ are fixed in the fit. Asymmetry and sigma are
the only two variables allowed to vary in the fit procedure. The
asymmetry variable is  a direct measure of the difference between the
parts of the shower below and above $X_{max}$. Sigma gives a measure
of the width of the shower. Asymmetry: $MF = 1.08$ and Sigma: $MF =
1.07$.

\vspace{0.3cm}

\underline{Skewness:} is the third momentum of the distribution and is
also a measurement of the asymmetry of the longitudinal
distribution. $MF = 2.13$

\vspace{0.3cm}

\underline{Kurtosis:} is the fourth momentum of the distribution and is
a combined measurement of the size of the peak and the tails. $MF =
1.69$

Other parameters were tested and rejected. For instance, we have fit a linear function in the range 
between $[X_{max} - 400,X_{max} - 100]$ and a second linear function in the range between 
$[X_{max} + 100,X_{max} + 400]$. The slope of these functions were taken as a measurement of 
the increase and decrease rate of the number of particles in the shower. However, no combination 
of the slopes has shown any capability to separate proton from iron showers.

\begin{figure}[]
\centerline{\includegraphics[width=13cm, angle=90]{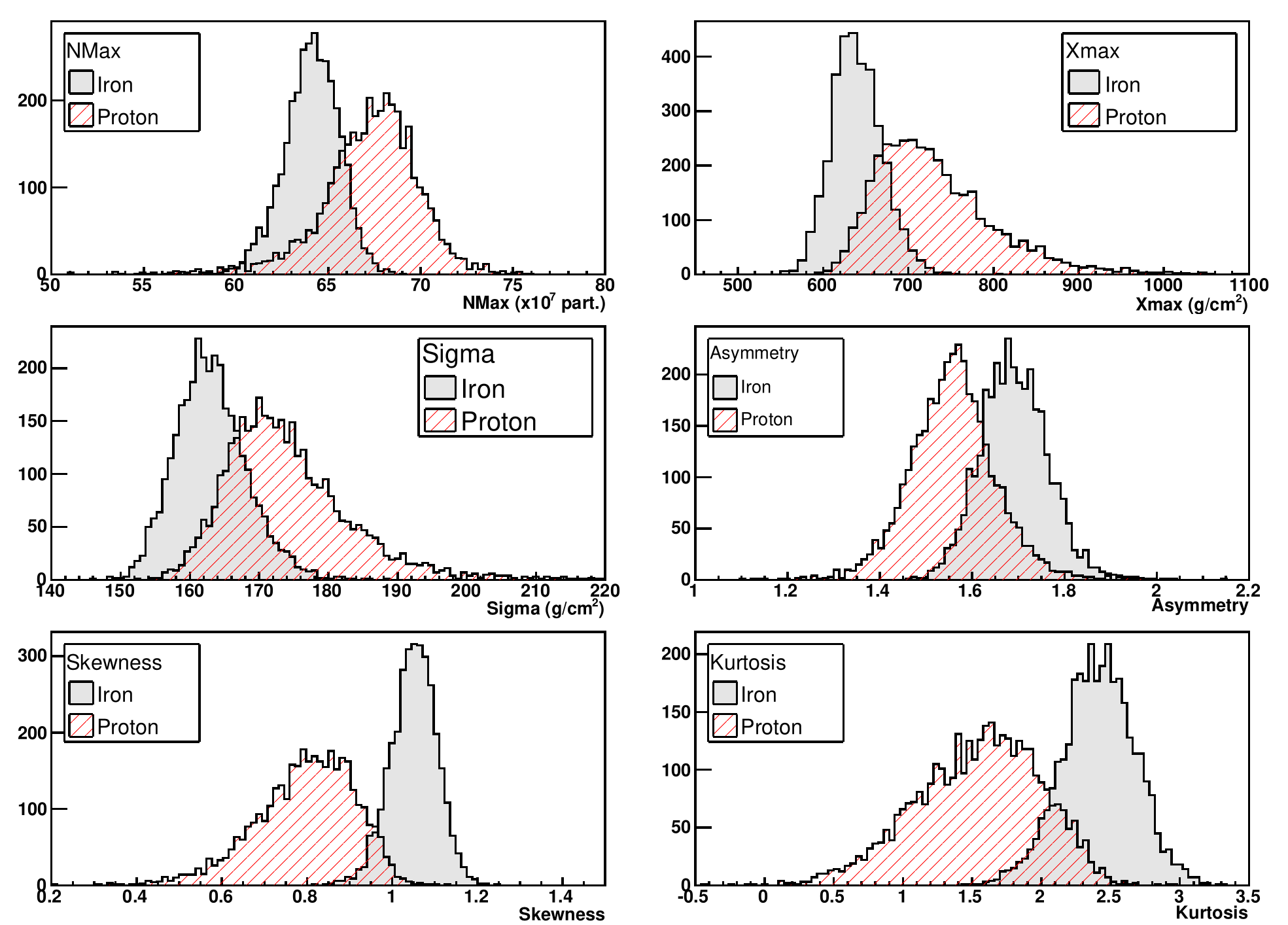}}
\caption{Distribution the $X_{max}$, $N_{max}$, sigma,
  asymmetry,skewness and kurtosis for proton and iron nuclei shower
  with primary energy $E_0 = 10^{18}$ eV.}
\label{fig:parameters}
\end{figure}

The distribution of the six parameters used in the composition studies proposed here are shown 
in figure \ref{fig:parameters} for proton and iron nuclei primaries.

Principal component analysis (PCA) was used to study the relevance of
each parameter to the determination of the shower overall features. PCA is
a linear transformation that rewrites a dataset into a new set of
variables that are uncorrelated. These new variables, called principal
components, are ordered by the amount of variance~\cite{bib:pca_01}. 

PCA has wide applications in astrophysics in analysis of spectra and galaxy surveys  
\cite{bib:Berial_06,bib:glaz_98} and in analysis of the Gamma Ray Bursts' spectra 
\cite{bib:bago_05}, parameter reduction in Dark Energy studies \cite{bib:lind_05}, finding 
systematic differences between stellar populations \cite{bib:ferr_06}, among others. 
In cosmic ray research, it has been used as a method to distinguish photon generated showers 
from hadronic showers applied to the secondary particle distributions at ground level produced 
by photons and protons showers \cite{bib:fale_04}. In another work, PCA was used 
in the data analysis of \v{C}erenkov Gamma Ray Telescopes \cite{bib:cont_06}.

\begin{figure}[]
\centerline{\includegraphics[width=7cm]{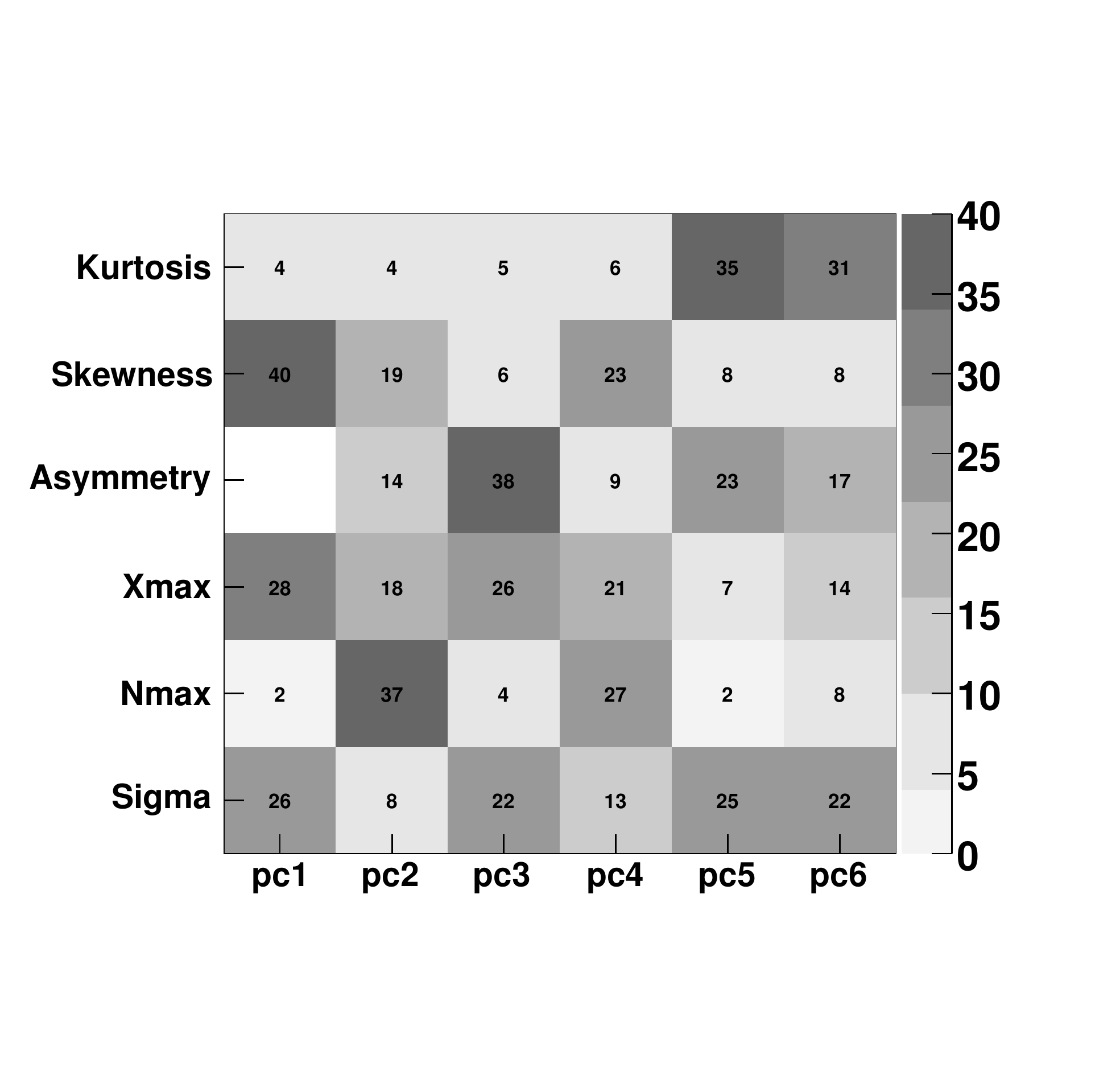}
\includegraphics[width=7cm]{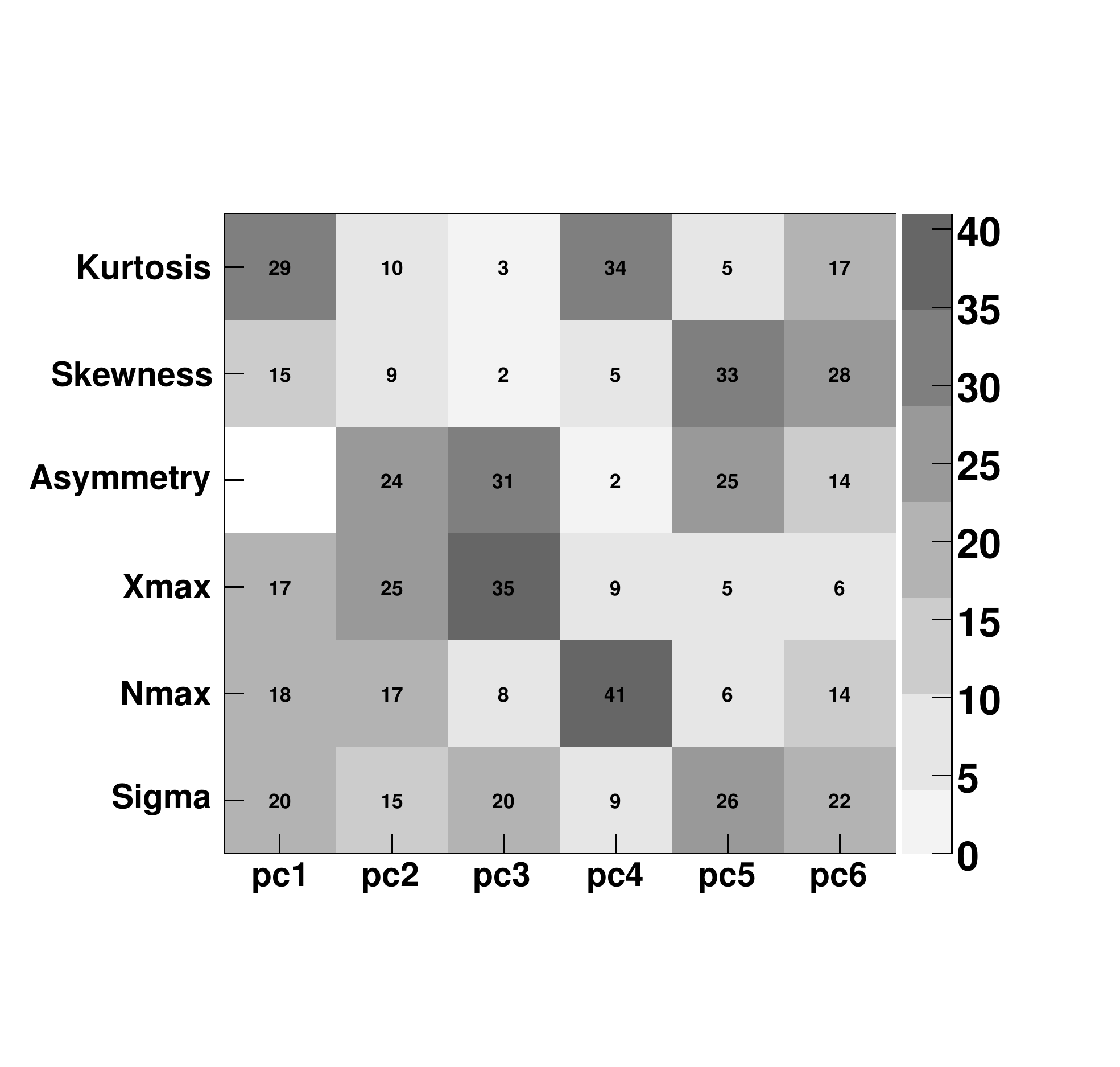}}
\caption{Weight of the parameters $X_{max}$, $N_{max}$, asymmetry,
  sigma, skewness and kurtosis in the PCA defined parameters
  f1, f2, f3, f4, f5 and f6 for proton (a) and iron(b) shower with energy 1 EeV. } 
\label{fig:weight}
\end{figure}

We have applied PCA for proton and iron showers and defined the eigenvectors
which characterizes their longitudinal profile. The first principal component is the eigenvector 
with the largest variance, related with the largest eigenvalue, the second principal component 
is associated with the second largest eigenvalue and so on. Figure \ref{fig:weight}a shows the 
contribution of each parameter to the principal component parameters when applied to the proton 
initiated showers. Figure \ref{fig:weight}b corresponds to the 
equivalent plot for the iron initiated showers. The weights shown in the figures were normalized 
to 100\%. It is interesting to note that the relative weight of the new parameters such as the 
skewness and the kurtosis is considerably different between the two data sets. 
Also, by comparing the variance of the principal components distributions from the two data 
sets, we have noted that some of the principal components show different profiles, indicating 
that indeed there are clear differences between the relative weights of the featured parameters 
to the overall shower determination. 
To take advantage of these differences for discriminating the two different data sets, we have 
applied a statistical method known as the linear discrimination analysis.

\section{\label{sec:lda} Linear discriminant analysis} 

Linear discriminant analysis (LDA) was used to search for the best combination of parameters 
to separate proton from iron nuclei showers and photon from hadronic primaries. 
LDA is a statistical discrimination method used to find a function of linear combinations of 
variables that maximizes the separation between two or more classes of objects or events. 
It accomplishes that by maximizing the ratio of the variability between different groups, determined through
the pooled (overall) covariance matrix, to the variability within each group (covariance matrix of each class)\cite{bib:lda}. 

\begin{figure}[]
\centerline{\includegraphics[width=7cm]{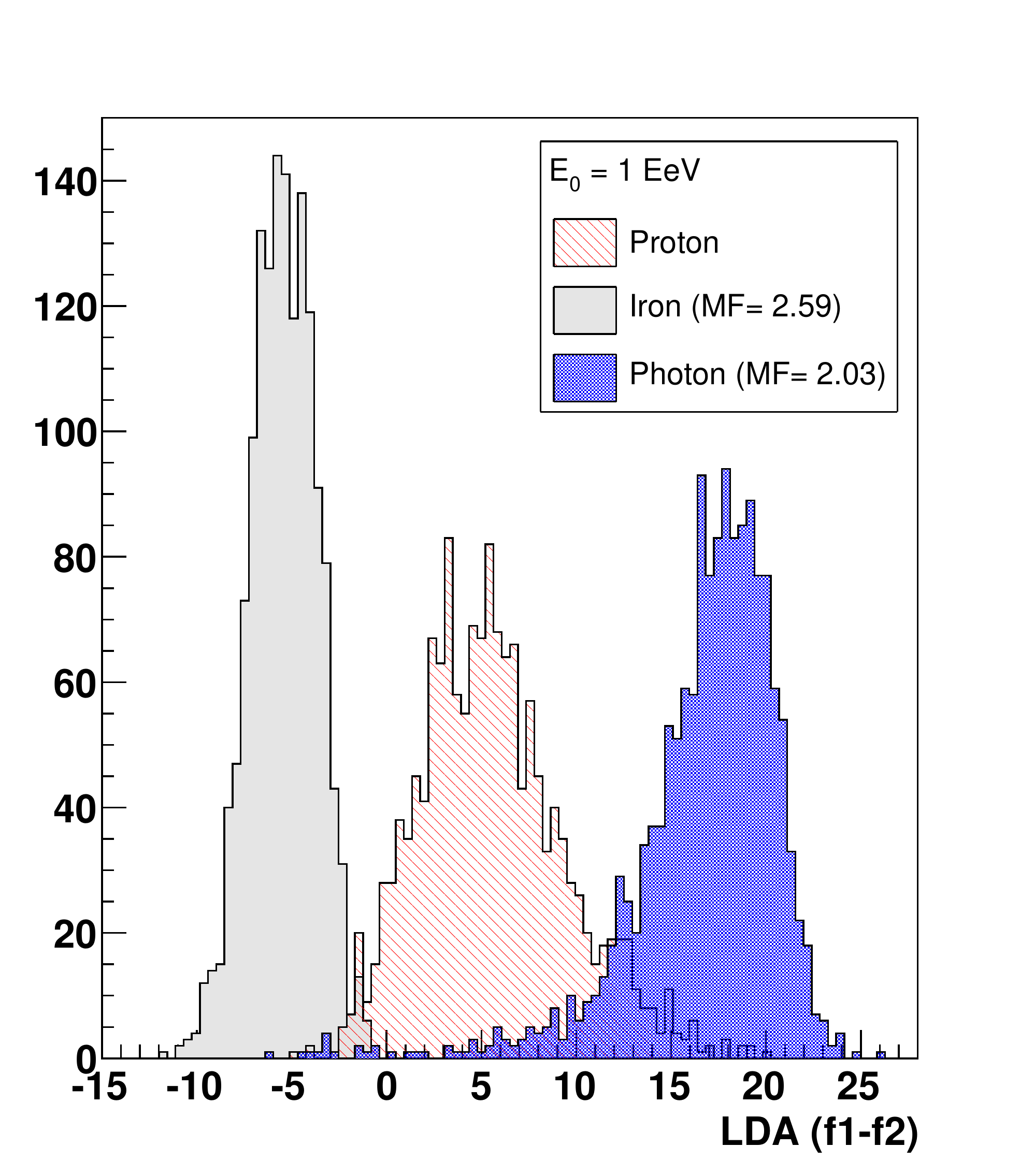}
\includegraphics[width=7cm]{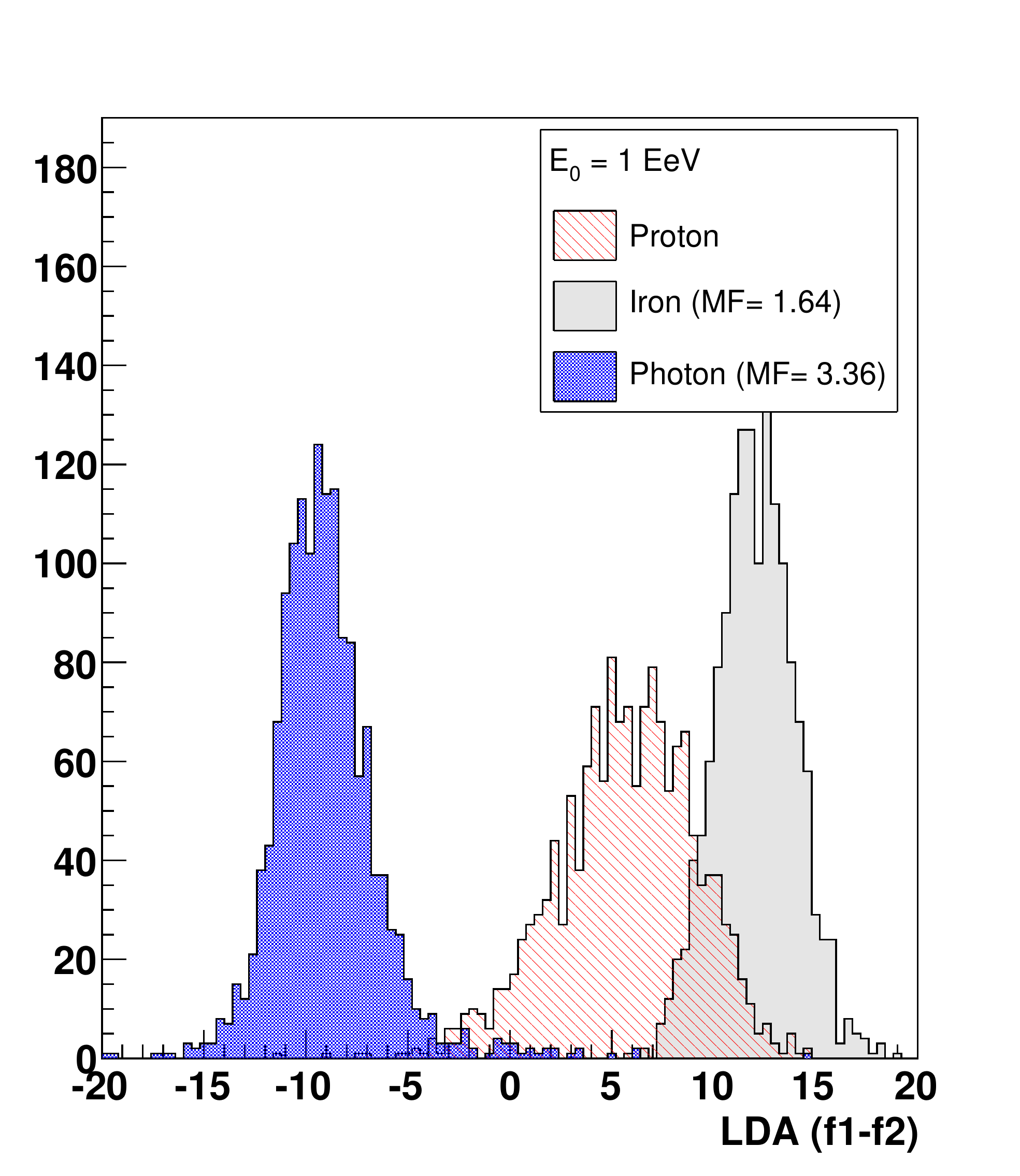}}
\caption{Distribution of LDA parameters ($f1-f2$) calculated for proton, iron and photon showers, using protons and irons as LDA training datasets in the left plot and using photon and hadrons as LDA training datasets in the right plot.}
\label{fig:pca:distri}
\end{figure}

In our analysis all six parameters presented in
figure~\ref{fig:parameters} were included in the LDA calculation.  
Training datasets for proton and iron showers were used to determine a
set of discriminant coefficient vectors. 
These discriminant coefficient vectors are then used to calculate the LDA variables $f1$ and 
$f2$ for each new data point, that was not in the original training dataset. The difference between the two 
LDA parameters ($f1-f2$) is expected to yield the best separation between the two populations. 
Applying this method to the proton and iron populations,
we have obtained a merit factor of 2.59. Including showers originated by photons in the 
same analysis keeping the discriminant coefficients that were calculated for proton and iron
a separation merit factor of 2.03 was achieved between photons and protons 
\ref{fig:pca:distri}a. To improve the photon-hadron discrimination, we repeated the procedure but using 
as training datasets one group of hadrons (50\%proton and 50\% iron showers) and another group of photons 
induced showers in which case, a merit factor of 3.36 was achieved. 
It is worthwhile to emphasize that $X_{max}$ alone has a merit factor equal to 1.20 for proton and 
iron showers and a merit factor of 1.38 between proton and photon distributions.

\begin{figure}[]
\centerline{\includegraphics[width=7cm, angle=90]{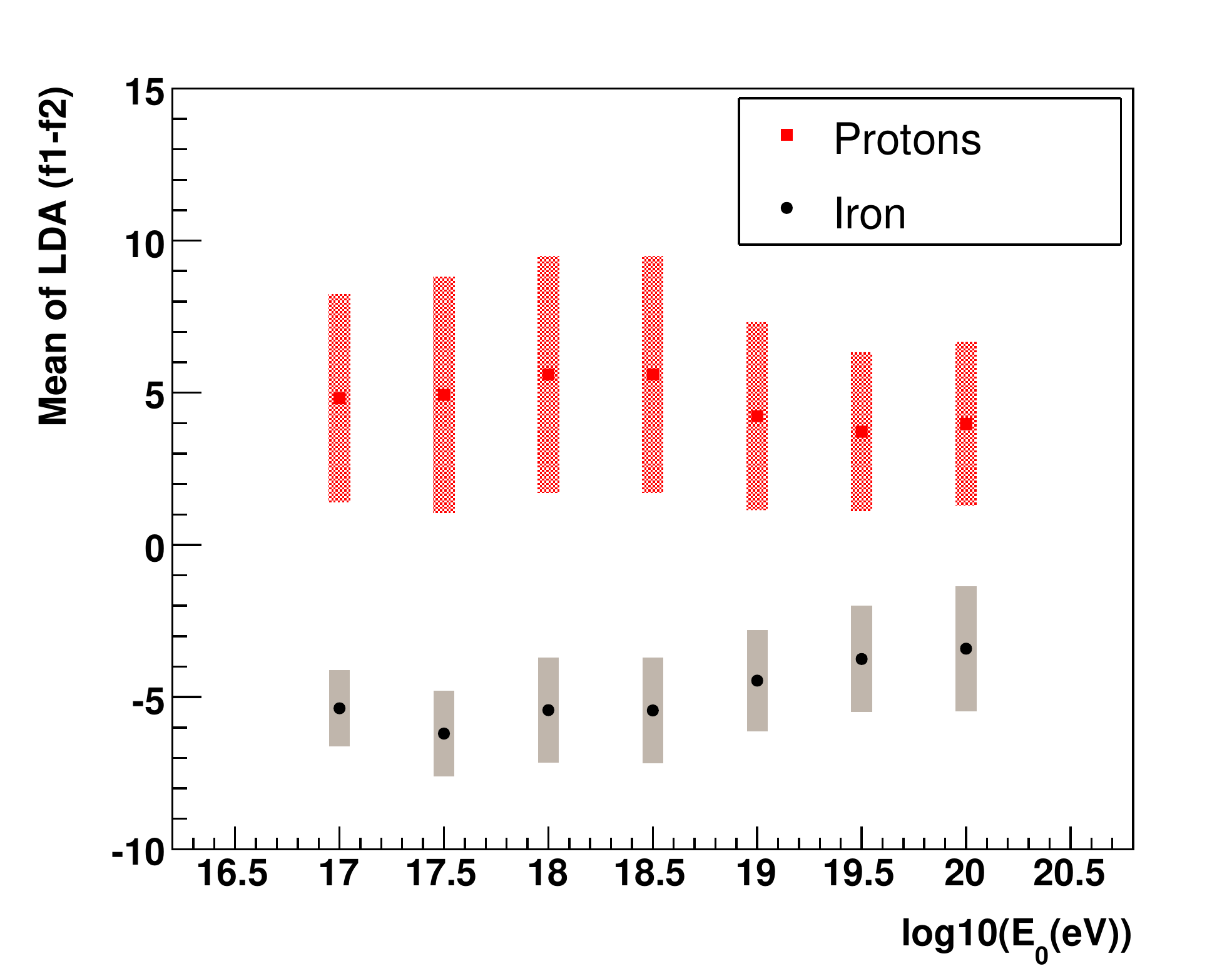}}
\caption{LDA parameter mean for protons and iron showers as a function of energy. 
Error bars represent the standard deviation of the distribution at each energy.}
\label{fig:lda:elongation}
\end{figure}

The capability of the LDA parameters was tested as a function of energy, where we have recalculated the 
dependent parameters coefficient for each energy. Figure \ref{fig:lda:elongation} shows the evolution 
of the LDA diagonal projection with energy calculated for proton and iron nuclei. This figure can be
compared to the $X_{max}$ elongation rate (see figure \ref{fig:elongation}). Figure \ref{fig:elongation} 
shows a overlap of the $X_{max}$ distribution along the entire energy range while figure
\ref{fig:lda:elongation} shows a clear separation of the LDA parameter along the same energy range.

\begin{figure}[]
\centerline{\includegraphics[width=7cm, angle=90]{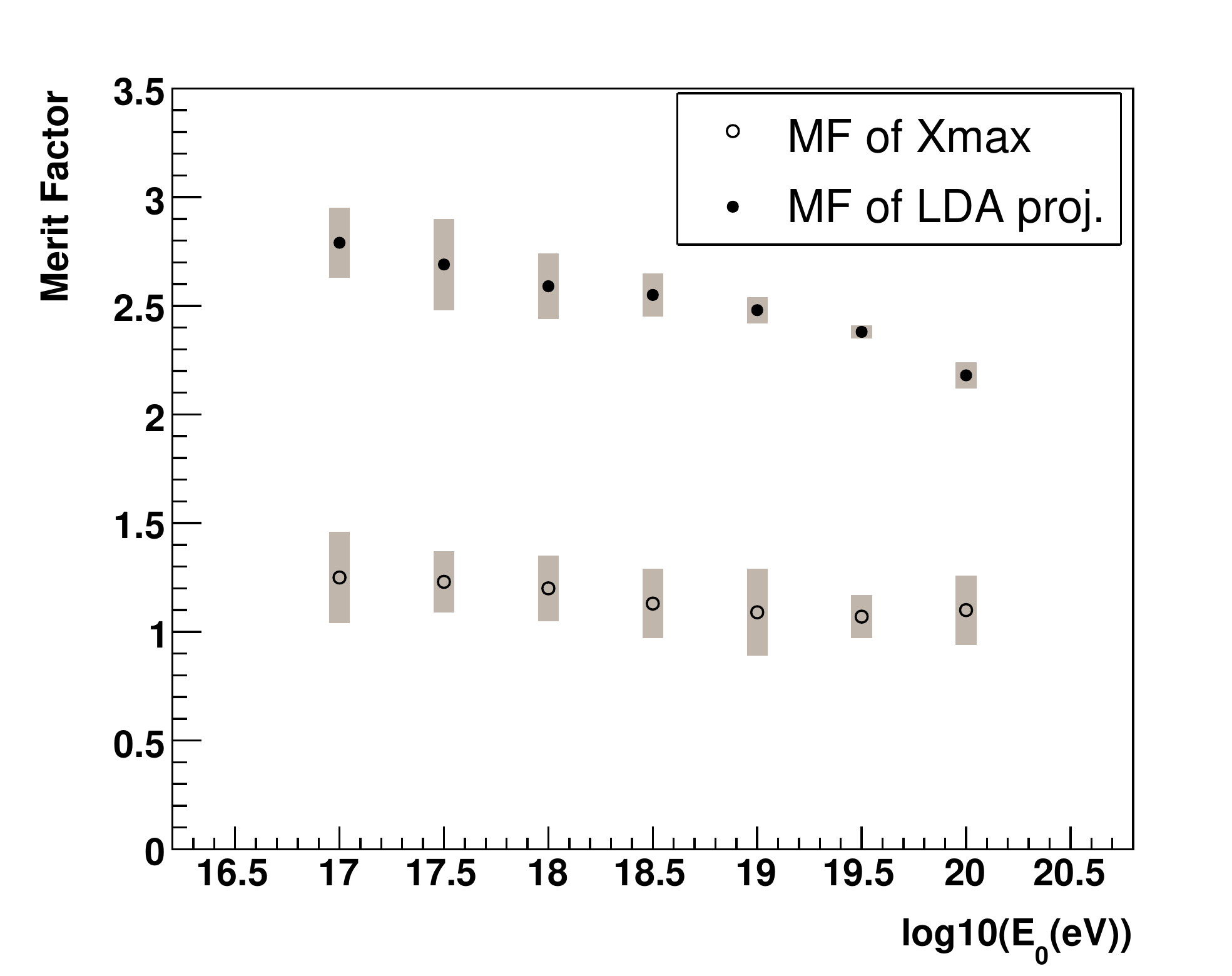}}
\caption{Merit factor between proton and iron distributions for the LDA
  parameter and $X_{max}$ as a function of energy.}
\label{fig:lda:energy}
\end{figure}

A comparison of both methods as a function of energy can be seen in figure \ref{fig:lda:energy} 
where we show the evolution of the merit factor for the LDA diagonal projection and for $X_{max}$ with 
energy.

\section{\label{sec:telescope} Simulation of the telescopes' field of view}

In order to test the effectiveness of the new analysis in a real situation with showers
detected by fluorescence telescope we have reduced the track length of the
showers used in our analysis and recalculated the discrimination
capability. We have also repeated the same LDA with a reduced part of
the longitudinal shower profile. Showers were limited to a maximum
length ranging from 400 g/cm$^2$ to 2000 g/cm$^2$ around its $X_{max}$.

The method presented here is based on the properties of the longitudinal profile and 
consequently the limited field of view of fluorescence telescopes is the main detection 
feature that affects the quality of the method. Detector fluctuations
and reconstruction 
uncertainties are expected to be a minor contribution to degrade the method capabilities. 
However, it is clear that the exact discrimination values of the new parameters must be 
calculated for each specific telescope configuration and analysis procedure.

\begin{figure}[]
\centerline{\includegraphics[width=7cm, angle=90]{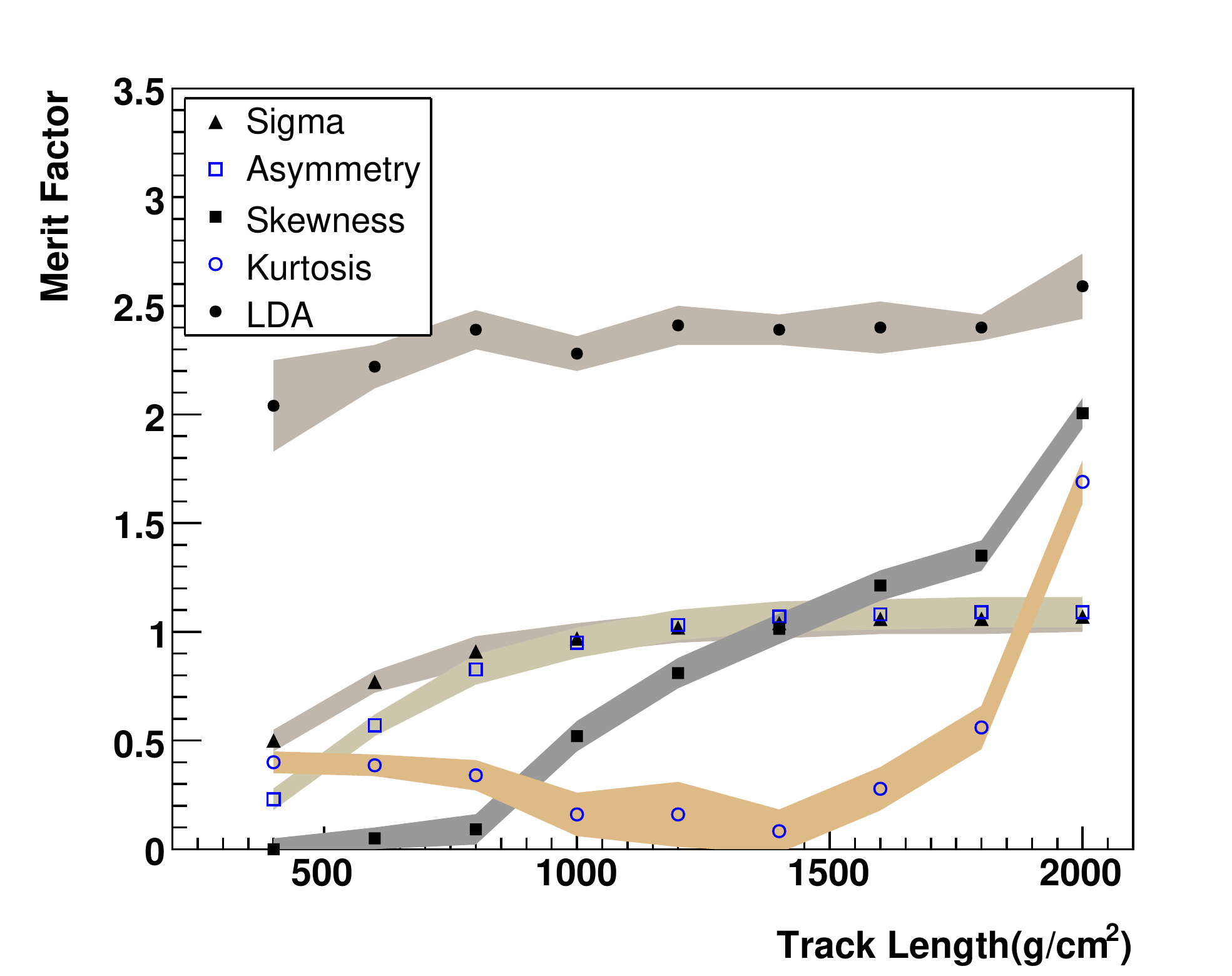}}
\caption{Merit factor between proton and iron distributions for shower
  parameters, sigma, asymmetry, skewness and kurtosis compared to the MF obtained from LDA 
as a function of profile slant depth.}
\label{fig:mf_all}
\end{figure}

Figure \ref{fig:mf_all} shows the merit factor of each parameter as a function of the 
total track length of the shower considered in our analysis. Calculations were done for 
showers with primary energy equals to $10^{18}$ eV but similar results were seen for all 
energies. As expected, kurtosis and skewness are efficient only when a large fraction of the 
shower is detected due to their dependence to the tails of the longitudinal distributions. 
Asymmetry and sigma remain good parameters even when the shower has 1000 g/cm$^2$, but it 
also degrades very fast for lower track length. Also shown are the merit factor calculated
using LDA, and which remained above 2 for all the entire range considered, showing that it 
yields better discrimination than considering each parameter independently.

\section{\label{sec:conclusion} Conclusion}

We studied different features of the cosmic ray shower longitudinal profile aiming to 
determine a better set of parameters that can improve primary particle identification. 
The relevance of each parameter to the overall profile was studied using a principal 
component analysis which showed that the relative weight of these additional parameters 
proposed in this analysis changed depending on the shower type.

The discrimination capability of the different parameters was compared using a merit factor
that measures how separate are the parameter distributions of the two different data sets. 
When evaluating the complete shower profile, we have determined that
some of the proposed parameters such as the distribution skewness and kurtosis show better 
discrimination capability than $X_{max}$ that is commonly used for composition analysis. 
However, these parameters loose very quickly its discrimination power when the track length of the
shower is reduced. 
To combine the separation capability of all the parameters the statistical method linear discrimination 
analysis was applied resulting in a new parameter with much better separation efficiency. 
With LDA, we were able to get separation merit factors above 2 sigma, 
between proton and iron initiated showers. For the separation between photon and hadron (proton+iron)
showers, LDA yields a separation merit factor of 3.3 sigma for shower energy of $10^{18}$ eV. 
Comparing to the composition analysis using the standard $X_{max}$, the LDA yields a
better separation efficiency, even when considering incomplete shower profiles. It is clear that
to have a better estimate of the applicability of this method to real data, a more complete 
simulation analysis considering a complete simulation of the detector response and atmospheric effects 
have to be considered.

In conclusion, we have shown that it is possible to improve composition studies of cosmic
ray showers by considering a more complete set of variables that describe better the shower
longitudinal profile. Further analysis with more complex and sophisticated statistical separation 
methods such as hierarchical clustering methods and neural network analysis can also further 
enhance this analysis.

\section{Acknowledgment}

This work was supported by the Brazilian science foundations FAPESP and CNPq to which we are 
grateful. We would also like to thank Prof. Dr. Carlos O. Escobar for the discussions and comments
that were essential to this work.

\newpage

\end{document}